\documentclass[10pt,conference]{IEEEtran}
\usepackage{cite}
\usepackage{amsmath,amssymb,amsfonts}
\usepackage{graphicx}
\usepackage{textcomp}
\usepackage{xcolor}
\usepackage[hyphens]{url}
\usepackage{hyperref}
\usepackage{booktabs}
\usepackage{tikz}

\usepackage[ruled,vlined]{algorithm2e}
\usepackage{multirow}
\usepackage{tabularx}
\usepackage{makecell}

\title{DiffusionSSD: Achieving Extreme Visual Compression with Dual-stage Diffusion and Cross-level Near-Data Processing}

\author{%
  \IEEEauthorblockN{Kuai Yu\IEEEauthorrefmark{1}\IEEEauthorrefmark{2},
    Wen Li\IEEEauthorrefmark{3},
    Shengwen Liang\IEEEauthorrefmark{1},
    Jie Zhang\IEEEauthorrefmark{4},
    Ying Wang\IEEEauthorrefmark{1},
    Huawei Li\IEEEauthorrefmark{1},\\
    Xin Yao\IEEEauthorrefmark{5},
    Renhai Chen\IEEEauthorrefmark{5},
    Gong Zhang\IEEEauthorrefmark{5}}
  \IEEEauthorblockA{\IEEEauthorrefmark{1}Institute of Computing Technology, Chinese Academy of Sciences\\
    \IEEEauthorrefmark{2}University of Chinese Academy of Sciences\\
    \IEEEauthorrefmark{3}School of Artificial Intelligence, Shanxi University\\
    \IEEEauthorrefmark{4}Peking University\\
    \IEEEauthorrefmark{5}Huawei Technologies Co., Ltd}
  \IEEEauthorblockA{\texttt{\{yukuai25,liangshengwen,wangying2009,lihuawei\}@ict.ac.cn}\\
    \texttt{liwen@sxu.edu.cn}, \texttt{jiez@pku.edu.cn}, \texttt{yao.xin1@huawei.com}\\
    \texttt{renhai.chen@tju.edu.cn}, \texttt{nicholas.zhang@huawei.com}}
}

\newcommand{\circnum}[1]{%
\tikz[baseline=(char.base)]{
\node[shape=circle, fill=black, inner sep=1pt] (char)
{\color{white}\small #1};}}

\begin{document}

\maketitle
\thispagestyle{plain}
\pagestyle{plain}

\begin{abstract}
The exponential growth of visual data overwhelms storage on resource-constrained edge devices, making compression essential. Traditional codecs suffer from inadequate rate-distortion performance, motivating the exploration of diffusion codecs. However, current diffusion codecs face critical limitations: (1) At the algorithm level, diffusion decoders preserve perceptual quality under extreme compression, but introduce substantial computation, making them far from efficient. (2) At the architecture level, in NPU-centered devices, neural codecs occupy the NPU, interfering with other AI applications. Meanwhile, the complex codec pipeline incurs long I/O paths, leading to additional overhead and energy cost.

To address these issues, we propose \textit{DiffusionSSD}, a cross-level near-data processing framework for extreme visual data compression on edge devices, using an algorithm-architecture co-design method. Specifically, (1) \textbf{at algorithm level}, we present \textit{Dual-stage Diffusion}, a lightweight solution tailored for diffusion codecs that employs patchification and single-step diffusion, reducing computation cost while preserving perceptual quality; (2) \textbf{at architecture level}, we introduce a \textit{Cross-level Near Data Processing} architecture that integrates the proposed computational universal flash storage (cUFS) and LPDDR-PIM, enabling codecs to be entirely offloaded to data side while reducing visual-data movement and avoiding SoC redesign; (3) \textbf{at system level}, combining both the algorithm and the underlying architecture, we correspondingly propose a \textit{Patch–Workload co-aware} scheduling scheme to enable more efficient system-level execution.
Experimental results show that \textit{Dual-stage Diffusion} maintains comparable perceptual quality while reducing computation cost by up to $92.7\%$. Compared with NPU+DRAM and near-data-processing design such as SmartSSD, cUFS+PIM improves coding throughput by $2.32\times$ and $2.91\times$ on average, respectively.

\end{abstract}

\section{Introduction}

The proliferation of high-resolution cameras and artificial intelligence-generated content (AIGC)\cite{aigc-survey} has democratized the acquisition and creation of high-fidelity visual data (e.g., 2K, 4K resolution), driving exponential growth in data volume that strains edge-device storage. While advances in 3D NAND technology\cite{3d-nand-survey} have enhanced storage density, they are still inadequate in meeting the increasing demands. For instance, emerging applications like high-frame-rate VR generate content at 17 GB/minute, amounting to 1TB/hour\cite{StorageHMD}, which rapidly exhausts the finite storage of edge platforms. %

\begin{figure}[t]
    \centering
    \includegraphics[width=\columnwidth]{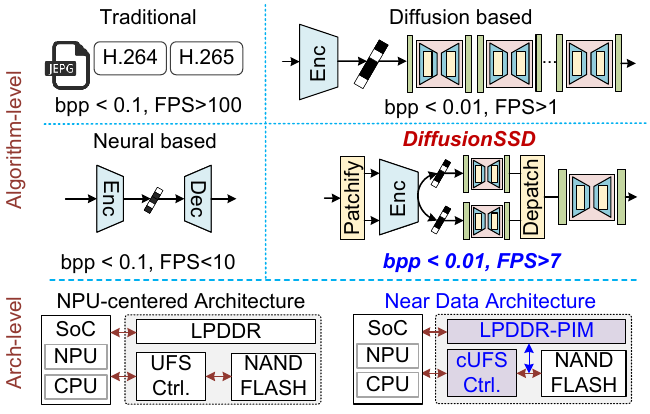}
    \caption{Neural codecs offer excellent compression quality but lower throughput. While diffusion-based codecs deliver better perceptual results, their frame rate usually drops below 1 fps. Our approach matches the perceptual quality of diffusion codecs while maintaining an acceptable frame rate.}
    \label{fig:intro}
\vspace{-10pt}
\end{figure}

Visual data compression is a promising solution to alleviate the growing
storage burden on edge devices. However, current visual data compression techniques employed in edge devices exhibit significant challenges. The slow pace of development and the limited compression ratios of both traditional lossless (e.g., Huffman coding\cite{huffman} and LZW\cite{lzw}) and lossy algorithms (e.g., JPEG\cite{jpeg2000} and H.264/265\cite{h264,h265}) drive researchers to explore neural compression methods. Recently, neural codec algorithms have shown a trend of surpassing traditional approaches in both reconstruction quality and compression efficiency.  Nevertheless, they remain far from practical deployment on resource-constrained edge devices. Specifically, 
\textit{\textbf{(1) at algorithm level}}, within conventional bits-per-pixel(bpp) ranges($bpp > 0.5$), many end-to-end neural codec works\cite{NVC-survey,NIC-survey,DVC,balle2018,neural-img-compression-1,neural-img-compression-2} achieve several-fold higher compression efficiency than traditional codecs at comparable reconstruction quality. 
In extreme bpp regimes($bpp < 0.5$), both traditional and neural codecs suffer from noticeable quality degradation. Latest studies employ diffusion-based approaches to preserve perceptual fidelity at high compression ratios, but the computation overhead of diffusion models leads to inefficient inference. These factors collectively erode the cost-effectiveness of diffusion solutions. 
\textit{\textbf{(2) at the architecture level}}, executing compute-bound diffusion codecs directly on the SoC NPU can occupy scarce accelerator resources for prolonged periods, interfering with concurrent intelligent applications~\cite{sereno}. Dedicated accelerators may alleviate such contention, but incur substantial area, power, and SoC redesign costs while offering limited reuse for other workloads. Moreover, codec execution still involves CPU-side entropy coding and frequent data movement, imposing additional host overhead.

To tackle these issues, we propose a \textit{Dual-stage Diffusion} and a \textit{Cross-level Near Data Processing} architecture, dubbed \textbf{\textit{DiffusionSSD}}, to enhance codec efficiency and system performance via a software-hardware co-design philosophy.

\textbf{\textit{At the algorithm level}}, to reduce the computation cost associated with high-resolution inputs, we apply a patchification preprocessing. The high-resolution image is partitioned into multiple smaller patches that are encoded and decoded independently, and later stitched back together. Moreover, to compensate for the loss caused by patching, we introduce a \textit{dual-stage diffusion} scheme: the first denoising stage performs local diffusion on each patch independently; the second denoising stage operates on the merged full-resolution image to restore global coherence. \textbf{\textit{At the architecture level}}, to prevent diffusion codecs from monopolizing NPU resources and interfering with other intelligent applications, we draw inspiration from compression-oriented SSDs designed for traditional codecs and propose a \textit{Computational Universal Flash Storage} (cUFS) architecture. By offloading diffusion codec to the data side, cUFS reduces NPU occupancy without redesigning the main SoC. However, due to the tight power and cost budgets of cUFS, as well as the absence of a sufficiently large DRAM on the UFS device, cUFS alone is insufficient to boost the diffusion model. To address this issue, we introduce a \textit{cross-level near data processing} architecture that integrates cUFS and LPDDR-PIM into a unified chip package using UFS-based multi-chip packaging (uMCP). A UCIe\cite{ucie} link directly connects cUFS and LPDDR-PIM, enabling them to collaboratively execute codec tasks while bypassing the SoC entirely, thereby eliminating unnecessary data movements. 
\textbf{\textit{At the system level}}, to enable efficient execution of \textit{Dual-stage Diffusion} on our architecture, we propose a \textit{patch–workload co-aware} scheduling to replace normal tile-based scheduling. In summary, our contributions include:

\noindent $\bullet$ We propose \textit{DiffusionSSD}, a cross-level near data processing framework for extreme visual data compression. To the best of our knowledge, \textit{DiffusionSSD} is the first near-data compression architecture designed for generative codecs. %

\noindent $\bullet$ We propose \textit{dual-stage diffusion}, a customized diffusion designed for a learning-based codec that adopts patchification and local-global context awareness to maintain high perceptual decoding quality while making the network more hardware-efficient.  

\noindent $\bullet$ We present a \textit{cross-level near data processing} architecture that integrates Computational Universal Flash Storage (cUFS) and LPDDR-PIM within a unified chip, avoiding SoC re-design and high NPU occupancy, while improving codec performance.

\noindent $\bullet$ A \textit{patch-workload co-aware} scheduling strategy is proposed to maximize the collaborative execution efficiency between cUFS and PIM.

\noindent $\bullet$ Evaluation shows that \textit{DiffusionSSD} achieves extreme compression ($<0.01$ bpp) with comparative visual quality over other diffusion-based methods. Compared to NPU+DRAM solutions, \textit{DiffusionSSD} improves throughput by $2.32\times$.

\section{Background}
\subsection{Neural Compression}

\begin{figure}[t]
    \centering
    \includegraphics[width=\columnwidth]{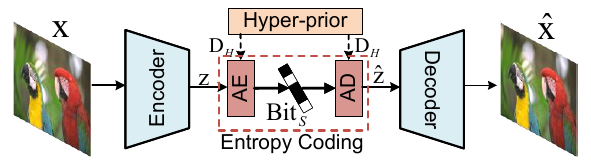}
    \caption{Encoder-Decoder Architecture of Neural Codec.}
    \label{fig:neural-codec-arch}
\end{figure}

\autoref{fig:neural-codec-arch} shows a neural compression scheme that combines traditional entropy models with encoder-decoder architectures. An encoder network $Enc$ converts input visual data $x$ into a latent representation $z$, which is then quantized and entropy-coded by an arithmetic encoder (AE) into a compact bitstream $Bit_S$.
Neural codecs enhance entropy modeling by introducing a hyper prior, an auxiliary latent that captures the spatial dependencies and statistical variance of $z$. Transmitted with side information, it helps to parameterize a learned distribution $D_H$ for more accurate entropy coding. During decoding, the bitstream $Bit_S$ is decoded to latent $\hat{z}$, which is then processed by a decoder $Dec$ to reconstruct the visual data $\hat{x}$. The parameters $Enc$ and $Dec$ are jointly optimized to minimize reconstruction error (distortion) and bitstream length (rate).

\textbf{Diffusion Codec}\cite{CDC, HDCompression, predictive-diffusion-1,predictive-diffusion-2} is an improvement over conventional neural codec. Diffusion-based methods replace the decoder with a generative diffusion process that synthesizes the output image conditioned on the entropy-decoded latent $\hat{z}$.
This generative formulation enables significantly better perceptual quality at low bitrates ($bpp < 0.1$). However, diffusion models introduce substantial computation overhead due to their iterative denoising procedure.

\subsection{Diffusion Model}
\textbf{Denoising Diffusion Models}\cite{ddpm, ddim, stable-diffusion} consist of a forward noising process and a reverse denoising process. In the forward process, gaussian noise is gradually added to the input image and a denoising net progressively denoises it in the reverse process:

\vspace{-1em}
\begin{equation}
\label{eq1}
q(\mathbf{x}_t | \mathbf{x}_{t-1}) = \mathcal{N}(\mathbf{x}_t; \sqrt{1 - \beta_t} \mathbf{x}_{t-1}, \beta_t \mathbf{I})
\end{equation}

\vspace{-1em}

\begin{equation}
\label{eq2}
p_\theta(\mathbf{x}_{t-1} | \mathbf{x}_t) = \mathcal{N}(\mathbf{x}_{t-1}; \mu_\theta(\mathbf{x}_t, t), \Sigma_\theta(\mathbf{x}_t, t))
\end{equation}

where $\mathbf{x}_t$ denotes the noisy image at timestep $t$, while $\beta_t \in (0,1)$ is a predefined parameter. The distribution $\mathcal{N}(\cdots; \mu, \Sigma)$ represents a multivariate gaussian distribution. In the forward process (Eq. \ref{eq1}), gaussian noise with variance $\beta_t$ is gradually injected into data, whereas in the reverse process (Eq.\ref{eq2}), the model parameterized by $\theta$ learns to predict the mean $\mu_\theta(\mathbf{x}_t, t)$ and covariance $\Sigma_\theta(\mathbf{x}_t, t)$ of the denoising distribution in order to recover $\mathbf{x}_{t-1}$ from $\mathbf{x}_t$.

\textbf{Few-step Diffusion}\cite{single-step} is an acceleration method that reduces the denoising process, lowering computation. Early few-step diffusion methods, such as progressive distillation\cite{Progressivedistillation} and distribution matching distillation\cite{distributionmatchingdistillation}, used multi-stage teacher-student frameworks that were computationally intensive and prone to losing fine details. Newer methods like Shortcut Models\cite{shortcut} and Fixed-Point Distillation\cite{Fixed-Point} simplify training with end-to-end objectives, enabling high-fidelity one-step generation. Few-step diffusion reduces latency, computational overhead, and memory bandwidth requirements.

\subsection{Mobile Storage \& Near Data Processing}
\begin{figure}[t]
    \centering
    \includegraphics[width=\columnwidth]{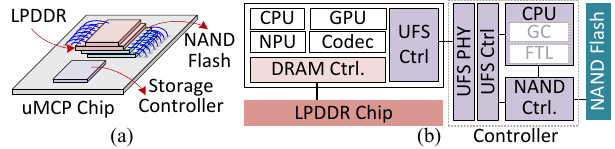}
    \caption{(a) uMCP overview. (b) The architecture of UFS chip.}
    \label{fig:ufs}
\end{figure}

\textbf{Universal Flash Storage (UFS)}\cite{UFS} is the successor to eMMC in modern edge devices, designed to provide substantially higher read/write bandwidth and lower latency. As shown in \autoref{fig:ufs}(b), a UFS chip integrates a UFS interface, embedded CPU, and NAND flash controller. The controller translates host logical block requests into physical NAND operations through the FTL. %
For edge devices with tight PCB area constraints, UFS is commonly packaged together with DRAM using a multi-chip package (MCP)\cite{uMCP}(\autoref{fig:ufs}(a)). However, MCP only provides physical co-packaging, where UFS and DRAM remain independent devices with separate interfaces and control paths. As a result, data transferred from UFS to DRAM must still traverse the host interconnect, UFS controller, and memory controller. Building on previous MCP integration efforts, this work enhances the high-bandwidth connectivity between UFS and DRAM within the MCP architecture, effectively minimizing data movement overhead.

\textbf{Computational Storage Device (CSD)} is a storage architecture that integrates computing capabilities directly into the storage system based on near data processing. The CSD couples compute units with storage components, placing them on the SSD controller\cite{cognitive-ssd, INSIDER, Inst-Attention, ECSSD, deepstore} or even within the flash die\cite{THRIFTY, OptimStore}. This design exploits the high internal bandwidth of NAND Flash and avoids data transfers over the PCIe or UFS interface to reduce data access overhead. Similarly, this work first identifies the limitations of neural codecs in NPU-centered architectures and then proposes a cUFS (computational UFS) architecture to offload the codec task from the NPU to the cUFS.

\textbf{Processing-in-Memory (PIM)} is another prominent instance of near-data processing that addresses the memory wall by embedding compute units within memory arrays or memory controllers, enabling partial or complete computation to be performed directly at the memory level. This approach substantially reduces data movement between memory accesses and computation, the corresponding energy cost, and improves system throughput. 
Leveraging PIM techniques not only prevents memory accesses from becoming the bottleneck in neural network execution but also provides additional compute capability. Through xPU–PIM collaborative execution, the system achieves higher overall throughput and a more balanced, reduced power distribution across edge devices.

\section{Motivation}

\begin{table}[t]
\centering
\caption{Performance of different codecs}
\begin{tabular}{lcc}
\toprule
\textbf{Algorithm} & \textbf{Enc.Thpt(FPS)} & \textbf{Dec.Thpt(FPS)} \\
\midrule
H.264   & 35.29 & 387.09 \\
Jpeg    & 143.69 & 142.64 \\
balle   & 21.75 & 36.51 \\
CDC-30  & 15.49 & 1.47 \\
\bottomrule
\end{tabular}
\label{motivation_fps}
\end{table}

\subsection{Diffusion Codec Optimization}
\begin{figure}[t]
    \centering
    \includegraphics[width=\columnwidth]{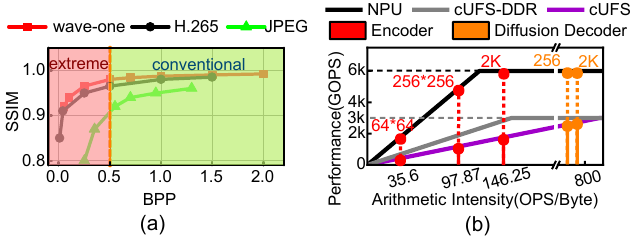}
    \caption{(a) Performance Collapse: For both traditional and neural approaches, when $bpp \leq 0.5$, the perceptual quality (SSIM) sharply falls. (b) Roofline analysis showing performance bottlenecks across different input resolutions, models, and hardware architectures.}
    \label{fig:motivation}
\end{figure}

Despite existing neural codecs having shown higher bit-rate efficiency, they still inherit two limitations: \textbf{First}, the \textit{performance collapse} problem. As shown in \autoref{fig:motivation}(a), in the conventional range, neural codecs\cite{waveone} outperform traditional codecs\cite{jpeg2000,h265}, as evidenced by their superior BPP-SSIM curve. However, at extreme ranges ($bpp < 0.5$), both neural and traditional codecs suffer significant degradation in reconstruction quality. \textbf{Second}, the \textit{computation costs}. We evaluate H.264 and JPEG on one Intel Xeon 8358P CPU (64 cores) and Balle’18\cite{balle2018}, CDC\cite{CDC} on one NVIDIA A100. As shown in \autoref{motivation_fps}, the decoding throughput of Balle’18 reaches only $25.5\%$ of JPEG’s throughput, highlighting the significant computational burden of neural codecs. 

To address the first limitation, many recent works adopt diffusion as a decoder \cite{CDC,DiffEIC,DiffVC} to improve perceptual quality (e.g., LPIPS/DISTS) at the extreme range. However, incorporating diffusion models further increases computational cost, stemming from the high-resolution input and the multi-step denoising process. %
A promising way to reduce the computational burden is to reduce the spatial size of the processed feature maps or shorten the denoising trajectory. The former is difficult to be applied to general tasks like Text2Image, which synthesize a globally coherent image. Fortunately, codec scenarios differ in that the input image acts as a reference signal for the encoded representation, allowing it to be partitioned into subregions for independent processing.

Although partitioning an image reduces the spatial cost, existing methods~\cite{CDC,DiffEIC,DiffVC} still suffer from the high overhead of multi-step denoising. This overhead arises from the conventional diffusion formulation, which initializes the reverse process from a high-noise state and uses the bitstream-derived upsampled feature only as an external condition, wasting the decoded structure and low-frequency content already preserved in the feature. We instead initialize it at a high signal-to-noise ratio (SNR) diffusion state. This enables the diffusion model to skip most of the reverse trajectory and avoid rediscovering information already encoded in the feature representation. This motivates our \textit{Dual-stage Diffusion}, a local--global scheme that performs one-step denoising on both patchified subregions and the stitched full image.

\subsection{Necessity of Cross-level NDP}

\label{NDP-motivation}
As shown in \autoref{motivation_fps}, conventional codecs generally provide fast decoding, matching the frequent-access pattern in practical use. Diffusion codecs exhibit the opposite behavior: their diffusion-based reconstruction makes decoding extremely slow (e.g., 1.47 FPS for CDC), causing prolonged contention for shared compute and memory resources and increasing user-perceived latency~\cite{sereno}. Inspired by prior SSD-based codec offloading designs~\cite{HA-CSD,zns-csd,balloon-zns,asic-based-compression-storage}, we explore offloading diffusion codecs to UFS to relieve the host, which introduces the following challenges:

\textbf{Architecture}. Diffusion codecs are predominantly compute-bound. As shown in~\autoref{fig:motivation}(b), we conduct a roofline analysis on an RK3588-like edge platform with a 6-TOPS NPU and 50-GB/s DRAM bandwidth~\cite{rk3588}. Although the encoder is memory-bound at low resolutions (64/256), both 2K encoding and diffusion decoding become compute-bound. Directly offloading such workloads to UFS is challenging because the compute capability that can be integrated under its power and area constraints is limited, while model weights and activations must be accessed from either host DRAM or flash. The former is constrained by the 4.0-GB/s UFS 4.0 interface, and the latter provides even lower bandwidth, particularly for writes.

Specifically, the bandwidth bottleneck arises from the access path between cUFS and the memory. Conventional cUFS accesses are limited by the 4.0 GB/s UFS interface. For the 3-TOPS cUFS configuration in ~\autoref{fig:motivation}(b), the roofline bound is \(P=\min(3000,4I)\) GOPS, where \(I\) denotes operations per byte transferred over this path. The workload points are in the bandwidth-bound region. Increasing the cUFS–DRAM bandwidth to 20 GB/s lowers the ridge point to 150 OP/Byte, motivating a direct DDR-based link that bypasses the host UFS path.
Since mobile DRAM and flash are commonly integrated outside the SoC through MCP technology and accessed via separate DDR and UFS interfaces, we combine UCIe with LPDDR-PIM~\cite{PIM-AI,samsung-pim}, which provides additional near-memory computation. Together, they form a cross-level near-data-processing architecture that alleviates both compute and bandwidth limitations without redesigning the host SoC.

\textbf{PIM Execution}. Although LPDDR-PIM augments the computing capability on the data side, existing designs typically support only a narrow set of operators, such as GEMV, and are less flexible than NPUs. Naively offloading only isolated supported operators would require frequent data transfers and host--PIM synchronization, whose overhead can offset or even dominate the computation benefits in edge inference. We therefore extend PIM support to common SIMD-friendly operators, including GEMM and activation functions, and enable it to execute precompiled sequences of consecutive supported operators without host intervention. This design reduces offloading boundaries and amortizes data-transfer and launch overheads.

\textbf{Scheduling Granularity}. After introducing cross-level NDP, the key question is no longer whether to offload computation, but at what granularity the workload should be scheduled. Directly adopting conventional layer/operator-level scheduling or fine-grained tensor mapping would fragment the execution of each input across heterogeneous computing units. The runtime must then track operator dependencies, coordinate bank accesses, preserve fine-grained data reuse, and frequently transfer intermediate results along the execution path. On resource-constrained edge systems, such coordination overhead can offset the benefits of heterogeneous execution.

\begin{figure*}[t]
    \centering
    \includegraphics[width=0.90\textwidth]{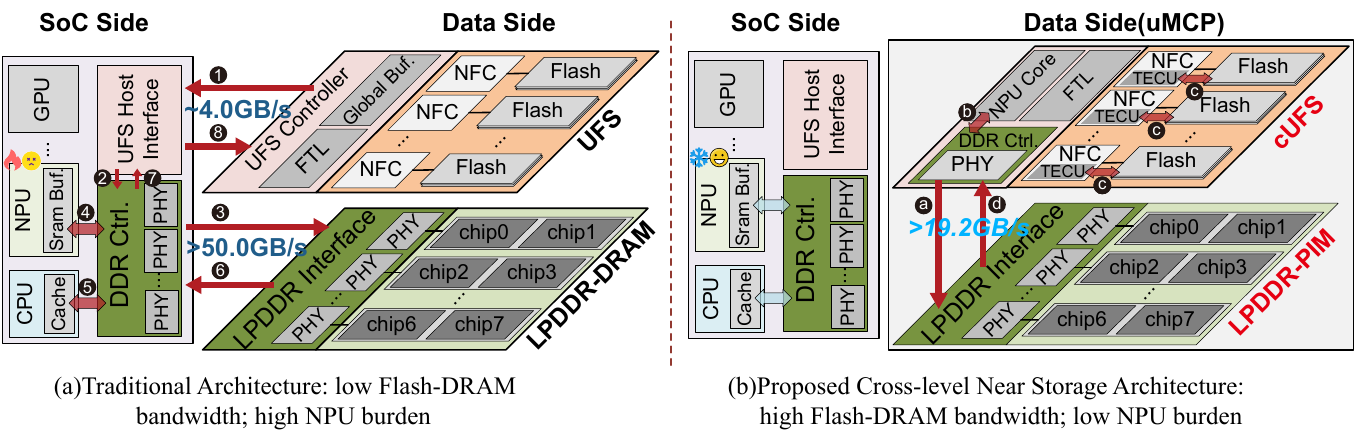}
    \caption{Cross-level Near-Data Processing Architecture.}
    \label{arch-design}
\end{figure*}

Fortunately, our algorithm-level optimization already partitions each image or video into independent patches. These patches follow the same computation graph and incur comparable execution costs, making them natural self-contained scheduling units. We therefore schedule workloads at the patch granularity, while leaving operator tiling to each computing device. This design avoids distributing fine-grained tiles across devices, and confines data transfers to a small number of predefined operator boundaries. This patch-level mapping reduces cross-device coordination and provides natural preemption points for latency-sensitive host workloads.

\section{Design}

\begin{figure}[t]
    \centering
    \includegraphics[width=0.90\columnwidth]{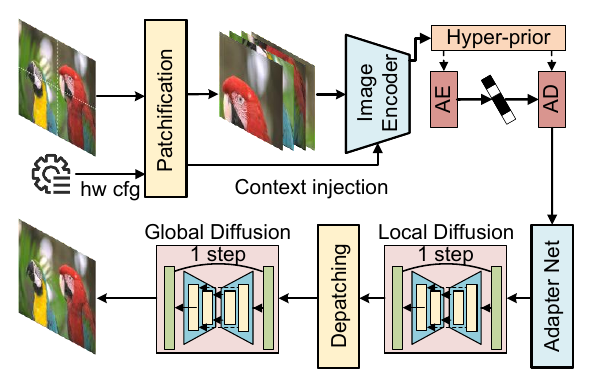}
    \caption{Diffusion codec pipeline with dual-stage diffusion.}
    \label{algorithm-design}
\end{figure}

\subsection{Dual-stage Diffusion}

Patchification is used in traditional codecs to manage high-resolution images by dividing them into low-resolution subregions, reducing memory and computational demands. However, this approach is complex for neural codecs, which depend on global features from full-resolution inputs and develop an entropy model for the entire image, as shown in \autoref{fig:neural-codec-arch}. This patch-based approach may disrupt long-range context and degrade quality. To overcome this, \textit{Dual-stage Diffusion} employs a local-global design: the local stage focuses on patch-level decoding to lessen spatial workload, while the global stage enhances coherence in the full-resolution output.

The second challenge is the high cost of multi-step diffusion adopted by existing diffusion codecs, which start the reverse process from a high-noise state, leaving the diffusion model to spend many unnecessary denoising steps. DsD instead uses the upsampled feature to initialize a high-SNR diffusion state.
Given the feature $\tilde{x}$, we sample a high-SNR diffusion state by:

\begin{equation}
\label{eq}
x_{t^*}=\sqrt{\bar{\alpha_{t^*}}}\tilde{x}+\sqrt{1-\bar{\alpha_{t^*}}}  \epsilon,\quad
\epsilon\sim\mathcal{N}(0,I)
\end{equation}
where $t^*$ is a fixed timestep selected from the high-SNR region of the diffusion schedule, so that $1-\bar{\alpha_{t^*}}$ remains small. 

The diffusion then predicts the reconstructed output through one denoising step:
\begin{equation}
\label{eq}
\hat{x}=D_{\theta}(x_{t^*},\tilde{x},t^*)
\end{equation}
where $\tilde{x}$ serves as the codec condition derived from the compressed representation. Under this formulation, the diffusion stage is not required to approximate the entire reverse trajectory.

The workflow of \textit{Dual-stage Diffusion} is illustrated in \autoref{algorithm-design}. The high-resolution input image is patchified into low-resolution tiles based on an optimal patch size. Each tile is processed through the encoder, entropy-encoded (AE), and stored as a bitstream. During decoding, the bitstream is entropy-decoded (AD), upsampled via an Adapter Network, and reconstructed by the local diffusion decoder (LD). After decoding all patches, a De-patching stage merges them into a full-resolution image. Lastly, global diffusion (GD) refinement compensates for context lost during patchification, delivering the final output image.

\subsection{Cross-level NDP Architecture}

\subsubsection{Architecture Overview}
\label{sec-arch-overview}
We propose a Cross-level Near-Data Processing (NDP) architecture using uMCP (UFS-based multi-chip packaging) to enable the diffusion codec offloading. As shown in \autoref{arch-design}(a), the typical mobile architecture includes a SoC with CPU, NPU, GPU, a memory controller (DDR Ctrl.), and UFS interface for storage, while the flash memory (co-packaged with the UFS controller) and LPDDR memory are located outside the SoC.

Taking the encoding procedure as an example: \textbf{First}, the visual data and model weights are fetched from flash and loaded into DRAM. During this process, the data is first read by the SoC-side UFS interface through \circnum{1}. It is then transferred through the DDR controller \circnum{2} and the LPDDR interface \circnum{3}, before being stored in DRAM. During this process, because UFS bandwidth ($\approx4.0$ GB/s) is significantly lower than LPDDR bandwidth ($>50$ GB/s), data moving between flash and DRAM is bottlenecked by the low UFS bandwidth. \textbf{Second}, the network inference(\autoref{fig:neural-codec-arch}) begins on the NPU-LPDDR subsystem \circnum{4}, and the generated latent features are stored in LPDDR memory. \textbf{Next}, the latent features enter the entropy coding stage(\autoref{fig:neural-codec-arch}), which involves a series of operations with complex control flow, bit-level manipulation, and frequent table lookups. Such operations are inherently control-intensive and exhibit limited data-level parallelism, therefore typically handled by the CPU or dedicated hardware accelerators for encoding and decoding. Modern neural compression pipelines typically offload entropy coding to the CPU+LPDDR\cite{balle2018,DVC,CDC} subsystem \circnum{5}. Finally, the encoded bitstream is written back to flash memory through the path \circnum{6} \circnum{7} \circnum{8}, completing the encoding pipeline. Overall, this workflow results in a long I/O path and places heavy dependence on the NPU, leading to substantial system overhead.

As shown in \autoref{arch-design}(b), we present the \textit{Cross-level NDP architecture}, which requires no redesign of the SoC and instead enhances the UFS and LPDDR subsystem. (1) \textit{Computational UFS} (cUFS) combines compute units with UFS storage by embedding a neural codec engine into the UFS. This engine integrates an NPU core and a flash-side \textit{Tensor Entropy Coding Unit} (TECU), enabling on-device inference while supporting entropy coding without SoC NPU and CPU involvement. (2) LPDDR-PIM is a processing-in-memory device customized for mobile platforms. Banks in LPDDR-PIM are equipped with processing units(PUs), which allows PIM to execute most diffusion codec layers. 

\begin{figure}[t]
    \centering
    \includegraphics[width=\columnwidth]{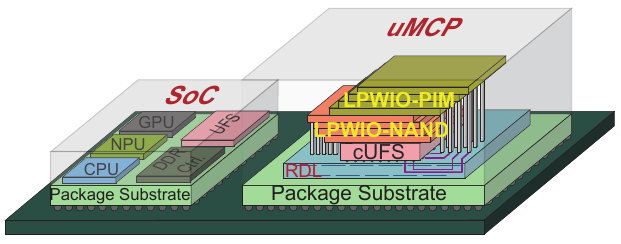}
    \caption{Cross-level NDP architecture in uMCP with vertically stacked NAND and PIM chips above cUFS, enabled by LPWIO interfaces and VFO-based advanced packaging}
    \label{uMCP-design}
\end{figure}

Building on uMCP architectures \cite{uMCP}, we integrate cUFS and PIM within a package to enable storage–memory collaborative computation. As shown in \autoref{uMCP-design}, the NAND and PIM dies are stacked above the cUFS controller. The cUFS-side DDR controller and PHY connect directly to the LPDDR interface of PIM, bypassing the host SoC and its UFS interface. Vertical interconnects implemented using VFO packaging \cite{kim2025vfo} route the stacked-die connections to the redistribution layer (RDL), whose wiring completes the connection to cUFS. The cUFS and PIM dies receive power through separate power connections routed from the package substrate.

Based on the Cross-level NDP architecture, the codec can be fully offloaded to cUFS-PIM. At the beginning of each run, model weights are transferred from flash through the UFS interface into LPDDR-PIM and remain resident for reuse across subsequent frames. Taking the encoding workflow as an example, the input image is directly fetched from the flash chips, and then transferred to LPDDR-PIM by path \circnum{a} ($>19.2$ GB/s). Next, cUFS and LPDDR-PIM collaboratively execute neural inference(NPU-LPDDR subsystem \circnum{b} and inside PIM). During the entropy-coding and write-back stages, latent features are first transferred back to cUFS via path \circnum{d} and encoded into bitstreams by the Tensor Entropy Coding Unit(TECU) inside the NAND Flash Controller (NFC), before being stored back to the flash chips through \circnum{c}.

\subsubsection{cUFS Design}
\begin{figure}[t]
    \centering
    \includegraphics[width=0.95\columnwidth]{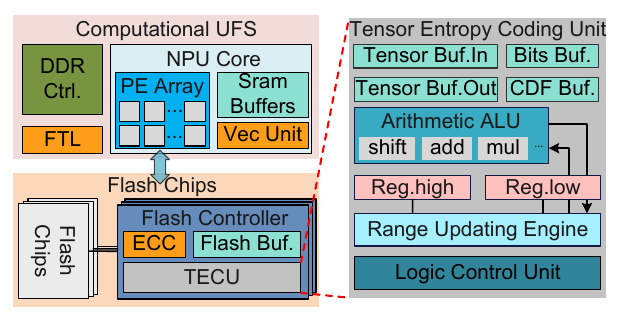}
    \caption{Computational UFS Details}
    \label{cUFS-design}
\end{figure}

The internal architecture of cUFS is illustrated in \autoref{cUFS-design}.
In our network, the downsampling network produces a low-resolution latent tensor that must be quantized and then entropy-coded into a bitstream before being stored in NAND flash. However, the NPU is not capable of handling entropy coding/decoding, and thus this task is typically offloaded to the CPU in conventional architectures. We observe that during encoding, latent features are entropy-coded before being stored into flash, while during decoding, the bitstream is fetched from flash and immediately entropy-decoded afterward. In both cases, entropy coding/decoding occurs directly adjacent to flash access. Therefore, placing entropy coding/decoding directly along the flash read/write path becomes a natural and efficient design choice.

We design the TECU to eliminate extra dependence on host CPU, as shown in \autoref{arch-design}(b). TECU comprises several submodules: an input tensor buffer (Tensor Buffer.In) for quantized features, a CDF table buffer (CDF Table Buf.) for probability distributions, a range updating engine with low and high registers (Reg.low and Reg.high) for interval updates, an optimized arithmetic ALU for necessary operations, a bitstream buffer (Bitstream Buf.) for output serialization, and an output tensor buffer (Tensor Buf.Out) for decoded symbols. The control unit coordinates CDF lookup, interval updates, and bitstream processing. During encoding, the input symbol indexes the CDF buffer to obtain the cumulative-probability boundaries used by the range updating engine to update the low/high registers. The control unit then sequences renormalization and bit emission. During decoding, the current code value and interval are used to locate the corresponding symbol in the CDF, followed by interval updates and bitstream refill. Thus, data-dependent branches and bit-level operations are handled locally by the control unit and arithmetic datapath. The CDF buffer holds codec-supplied probability tables, allowing different distributions to be used without changing the arithmetic-coding datapath. Compatible model updates require loading new weights and the corresponding CDF tables without hardware changes. Entropy-coding schemes unsupported by TECU can be executed on the host CPU.

\subsubsection{LPDDR-PIM}

\label{lpddr-pim}

We replace conventional LPDDR-DRAM with LPDDR-PIM to complement the limited compute capability of cUFS for offloading, and further augment PIM with general SIMD-style support. Unlike existing PIM designs that primarily accelerate GEMV operations, our design executes consecutive supported operators, reducing kernel-launch overhead and intermediate data movement. This capability enables practical offloading of diffusion codecs, while also providing energy-efficiency benefits through near-memory execution.

\begin{figure}[t]
    \centering
    \includegraphics[width=\columnwidth]{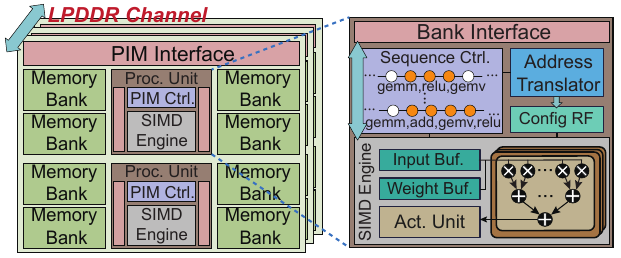}
    \caption{LPDDR-PIM design}
    \label{PIM-design}
\end{figure}

\autoref{PIM-design} shows the microarchitecture of LPDDR-PIM. In each PIM chip, one processing unit (PU) is shared by every four DRAM banks and accesses their internal bandwidth for near-bank processing. The operand and activation tiles of a patch can be striped across these four banks, but each patch is exclusively assigned to and processed by their shared PU. Each PU mainly consists of a PIM controller for coordinating consecutive operator execution and a SIMD engine that provides general SIMD-style compute capability. Specifically, the sequence controller within the PIM controller is configured before task execution to manage the precompiled operator sequence. At runtime, the address translator resolves the locations of the assigned patch data across the associated banks. 
The SIMD engine integrates multiple SIMD MAC lanes and a lightweight activation unit for element-wise operations.

\textbf{PIM-Supported Operators}. Our PIM mapping is selective. Since edge visual compression typically processes one frame at a time, fully connected layers are executed as GEMV-like kernels. We also support regular convolutions that can be executed as tiled matrix operations, together with element-wise operations implemented by the activation unit. Irregular convolutions and other unsupported operators are assigned to cUFS. These supported operators define the candidate operator segments considered by our scheduler.

\subsubsection{Memory Synchronization}
In our architecture, the SoC memory controller and cUFS-side DDR controller can both access LPDDR-PIM, which may lead to memory synchronization issues. To address this, we design a two-level synchronization mechanism: a hardware ownership arbiter and a system staging buffer protocol.

First, the LPDDR-PIM channel is managed by an ownership arbiter that toggles between SoC-owned and cUFS-owned modes. In SoC-owned mode, the SoC memory controller exclusively manages the channel like normal DRAM, while the cUFS-side DDR controller can only prepare data or submit requests. In cUFS-owned mode, the cUFS-side DDR controller takes over, handling cUFS-PIM transfers and related computations. This setup ensures that only one controller manages the LPDDR-PIM channel at a time.

Second, at the system level, we allocate a pinned staging buffer for data consistency. In cUFS-owned mode, the cUFS DDR controller writes to the staging buffer, keeping it invisible to SoC accesses until the transfer completes. Afterward, cache invalidation is performed, and the region is exposed as read-only to the SoC/NPU. In SoC-owned mode, the SoC writes to the buffer, performs cache clean, sealing the region as read-only for the cUFS DDR controller. This protocol is exclusive to the staging buffer used by the direct link; other memory regions are managed conventionally by the SoC memory controller. Cache maintenance occurs at host–cUFS data handoffs rather than at each layer. The staging-buffer allocation remains fixed, and its access permissions remain unchanged during a codec execution phase. Thus, invalidation is required when cUFS-produced data is handed back to the host, rather than repeatedly during internal cUFS–PIM execution.

\subsection{Patch-Workload Co-Aware Scheduling}
\label{sec-scheduling}

According to the LPDDR-PIM design in Section~\ref{lpddr-pim}, each PU can execute consecutive sequences of selected FC layers, regular patch-local tiled convolutions, and activation operators. However, unsupported operators involving complex reductions or irregular data rearrangements, such as \textit{LayerNorm} and \textit{PixelShuffle}~\cite{pixelshuffle}, must still be executed on cUFS. 

Conventional layer/operator-level scheduling or fine-grained tile mapping is ill-suited to this heterogeneous execution model. It fragments patch execution across devices and requires fine-grained dependency tracking.
More importantly, such local scheduling decisions cannot jointly account for operator-segment boundaries, patch allocation, and workload characteristics, making it difficult to fully exploit the cross-level NDP architecture.

\begin{figure}[t]
    \centering
    \includegraphics[width=0.95\columnwidth]{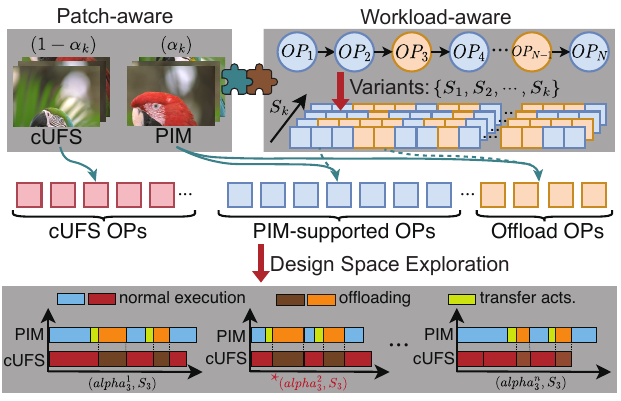}
    \caption{Patch-Workload Co-aware Scheduling.}
    \label{scheduling}
\end{figure}

\begin{figure*}[t]
    \centering
    \includegraphics[width=0.90\textwidth]{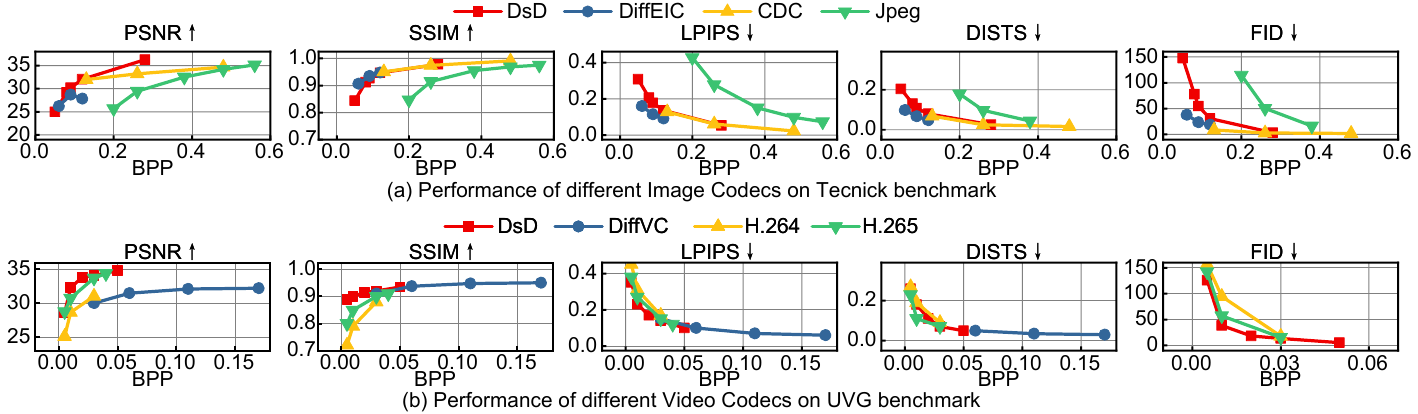}
    \caption{Algorithm performance of different tasks and benchmarks.}
    \label{alg-perform}
\end{figure*}

We propose the \textit{Patch-Workload Co-Aware Scheduling} strategy, as shown in \autoref{scheduling} that fully leverages the patch-independent property introduced by our dual-stage diffusion.

First, we begin by analyzing the neural codec network, such as encoding or decoding network parts, represented as an operator sequence:
\[
\mathcal{L}=\{Op_1,Op_2,\dots,Op_N\}
\]
Each layer is classified as either PIM-supported or PIM-unsupported. We denote all unsupported operators as $\mathcal{U}$. Given the location of operators in $\mathcal{U}$, we randomly generate a set of segment-level execution variants:
\[
\mathcal{S} = \{S_1, S_2, \dots, S_K\}
\]
where each $S_k$ is defined as:
\[
S_k = (P_0^{(k)}, C_0^{(k)}, P_1^{(k)}, C_1^{(k)}, \dots, P_{M_k}^{(k)})
\]
with $P_i^{(k)}$ and $C_i^{(k)}$ denoting maximal contiguous on-PIM and on-cUFS operator sequences, respectively. The segmentation implicitly reduces data movements between cUFS and PIM by grouping multiple PIM-supported operators into a sequence, avoiding fine-grain and fragmented round-trip behavior.

Second, we analytically model the patch-aware cost under the segmentation strategy $(S_k)$.
Assume that a total of $B$ patches are to be encoded or decoded. A fraction $\alpha_k$ ($0 < \alpha_k < 1$) is managed by the PIM, while the remaining $1 - \alpha_k$ is handled by cUFS. Due to its operator limitation, PIM cannot execute the entire workload and thus requires cooperative execution with cUFS. In contrast, cUFS can independently process the remaining $1 - \alpha_k$ portion end-to-end.

We introduce a device-level timing estimation function 
$\text{EstTime}(\cdot)$ to estimate the execution time of an operator segment on either PIM or cUFS. 
For a device $d\in\{\mathrm{PIM},\mathrm{cUFS}\}$, we simply model:
\[
\text{EstTime}(\text{Cfg}_d,o)
=
\max\left(
\frac{\mathrm{Ops}(o)}{\eta_d^{\mathrm{comp}}P_d},
\frac{\mathrm{Bytes}(o)}{\eta_d^{\mathrm{mem}}B_d}
\right)
+
T_{\mathrm{stall}}^{d}(o),
\]
where $P_d$ and $B_d$ denote peak compute throughput and effective memory bandwidth, $\eta_d^{\mathrm{comp}}$ and $\eta_d^{\mathrm{mem}}$ capture efficiency loss, and $T_{\mathrm{stall}}^{d}(o)$ is the non-overlapped stall time obtained from an event scheduler, covering command queuing, bank/channel contention, memory conflicts, and device-level synchronization. 
For scheduling-time estimation, we set $\eta_{\mathrm{comp}}^d$ as an empirical value accounting for dataflow and tiling inefficiencies. We derive $\eta_{\mathrm{mem}}^d$ as the ratio of effective bandwidth obtained from memory trace analysis to the reference bandwidth $B_d$.

For PIM-executed segment $P_i^{(k)}$, the compute time is estimated as:
$$
T_{\text{PIM},i}^{(k)} = \text{EstTime}(\text{Cfg}_{\text{PIM}}, P_i^{(k)})
$$

When the execution of $C_i^{(k)}$ is required, LPDDR-PIM must offload the output activations of 
$P_i^{(k)}$ (size $A_i$). 
Given the effective PIM-cUFS bandwidth $B_{\text{off}}$, 
the transfer time is expressed as:
$$
T_{\text{off},i}^{(k)} = {A_i}/{B_{\text{off}}}
$$

Once activations arrive at cUFS, the cUFS executes $C_i^{(k)}$ immediately, 
pausing its ongoing local execution.

Such execution stalling is captured by:
$$
T_{\text{blk},i}^{(k)}
= \beta_{\text{blk}} \cdot 
\text{EstTime}(\text{Cfg}_{\text{cUFS}}, C_i^{(k)}),
$$
where $\beta_{\text{blk}}>1$ models the runtime blocking and scheduling overhead 
caused by handling these intermediate offloads.

Aggregating across all segments in $S_k$, 
the per-patch estimated time on PIM and cUFS is:
$$T_{\text{PIM}}^{(k)} 
= \sum_{i} \Big(T_{\text{PIM},i}^{(k)} + T_{\text{off},i}^{(k)} \Big)$$

$$T_{\text{cUFS,blk}}^{(k)} 
= \sum_{i} T_{\text{blk},i}^{(k)}$$

Meanwhile, for the remaining $1-\alpha_k$ fraction, their per-patch cost is:
$$
T_{\text{cUFS,base}} 
= \text{EstTime}(\text{Cfg}_{\text{cUFS}}, \mathcal{L})
$$
where $\mathcal{L}$ denotes the full network.

The total steady-state busy time on PIM and cUFS becomes:
$$
T_{\text{PIM,tot}} 
= \text{Overlap}(\alpha_k B, T_{\text{PIM}}^{(k)})
$$
$$
T_{\text{cUFS,tot}} 
= \alpha_k B \cdot T_{\text{cUFS,blk}}^{(k)}
+ \text{Overlap}((1-\alpha_k) B, T_{\text{cUFS,base}})
$$
where $\text{Overlap}(\cdot)$ captures the overlapped time via multi-patch pipelining.

To maximize system throughput, our optimization objective:
$$
\max_{\alpha_k}
\frac{B}{\max\left(
T_{\text{PIM,tot}},
T_{\text{cUFS,tot}}
\right)}
$$

Finally, by enumerating all segmentation strategies 
$\{S_1, S_2, \dots, S_K\}$ and solving the optimal patch ratio $\alpha_k$ for each strategy as formulated above, the overall optimal configuration $(S_{k^*}, \alpha_{k^*})$ is obtained. %
In this case, patches can be pre-placed in LPDDR-PIM, highlighting opportunities for design space exploration in compute allocation.

\section{Architecture Feasibility}

\begin{figure}[t]
    \centering
    \includegraphics[width=0.95\columnwidth]{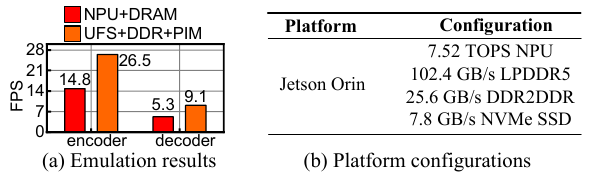}
    \caption{The emulation results and platform configurations.}
    \label{realsim}
\end{figure}

To demonstrate the feasibility of the proposed cross-level NDP architecture, we first build a prototype system based on NVIDIA Jetson Orin and perform end-to-end diffusion codec execution through hardware emulation. To emulate the proposed uMCP architecture, we partition the LPDDR5 DRAM on Orin into two regions at the DDR-channel granularity, representing the memory resources associated with the UFS and DRAM, respectively. We use the DDR-to-DDR to emulate the UCIe link. Meanwhile, the DLA and its associated DRAM are used to emulate cUFS. We emulate PIM in software. Specifically, we first precompute the intermediate results of all patches according to the configured PIM architecture. During emulation, for each patch assigned to PIM, we use the precomputed fixed latency and memory access time. The configuration of the prototype system is listed in \autoref{realsim}(b).

\begin{table}[t]
\centering
\caption{Architecture Simulator Configuration}
\begin{tabular}{|c|ccc|}
\hline
Architecture   & \multicolumn{1}{c|}{NPU}      & \multicolumn{1}{c|}{cUFS}    & cUFS+DDR \\ \hline
Peak OPs     & \multicolumn{1}{c|}{6 TOPS}    & \multicolumn{1}{c|}{4 TOPS}   & 4 TOPS    \\ \hline
Sram Size      & \multicolumn{3}{c|}{act.256KB, wt.64KB, out.256KB}                      \\ \hline
Peak Bandwidth  & \multicolumn{1}{c|}{51.2GB/s} & \multicolumn{1}{c|}{4.0GB/s} & 19.2GB/s \\ \hline
Entropy Coding & \multicolumn{3}{c|}{60fps@2K input}                                     \\ \hline

\multirow{4}{*}{LPDDR-PIM} &
  \multicolumn{3}{c|}{\multirow{4}{*}{\begin{tabular}[c]{@{}c@{}}
  4 channels, 2 PIM chips/channel\\
  2.4 TOPS/channel\\
  BW\_channel=19.2GB/s\\
  BW\_onchip=102.4GB/s
  \end{tabular}}} \\
               & \multicolumn{3}{c|}{}                                                   \\
               & \multicolumn{3}{c|}{}                                                   \\
               & \multicolumn{3}{c|}{}                                                   \\ \hline

\multirow{4}{*}{Flash Memory} &
  \multicolumn{3}{c|}{\multirow{4}{*}{\begin{tabular}[c]{@{}c@{}}
  4 channels, 2 Flash Chips/channel\\
  1200MT/s@8bit,page\_size=16KB,\\
  t\_rp=20us,t\_wp=500us
  \end{tabular}}} \\
               & \multicolumn{3}{c|}{}                                                   \\
               & \multicolumn{3}{c|}{}                                                   \\
               & \multicolumn{3}{c|}{}                                                   \\ \hline

\multirow{5}{*}{Power Modeling} &
  \multicolumn{3}{c|}{\multirow{5}{*}{\begin{tabular}[c]{@{}c@{}}
  NPU core 0.2pJ/op, LP-DRAM 50pJ/byte;\\
  PIM external 50pJ/byte, internal 5pJ/byte,\\
  PIM compute 0.4pJ/op; TECU 40mW\\
  UFS read 1.0nJ/byte, write 2.0nJ/byte;\\
  PHY read 130mW, write 230mW
  \end{tabular}}} \\
               & \multicolumn{3}{c|}{}                                                   \\
               & \multicolumn{3}{c|}{}                                                   \\
               & \multicolumn{3}{c|}{}                                                   \\
               & \multicolumn{3}{c|}{}                                                   \\ \hline

\end{tabular}

\label{cUFS-cfg}
\end{table}

\autoref{realsim}(a) shows that on the Jetson Orin platform, our architecture improves the throughput of the encoder and decoder by 1.79$\times$ and 1.72$\times$, respectively. 
However, the prototype is constrained by the available Orin hardware: cUFS is emulated using the DLA, the direct link through DDR-to-DDR transfers, and PIM in software. These components cannot reproduce the physical organization and resource configuration of the target architecture. We therefore develop a SCALESim-style simulator\cite{scalesim} to evaluate the target cUFS–PIM system.
The configurations are listed in \autoref{cUFS-cfg}. We use the RK3588\cite{rk3588} specification as a baseline for mobile platforms. The UFS and DDR controllers are based on UFS 4.0 and a single DDR channel.
The simulated configuration represents an RK3588-like edge platform and differs from the Orin prototype in compute capability and memory bandwidth. The speedups in each setup are measured relative to its corresponding baseline.

We build a system-level energy/power model (\autoref{cUFS-cfg}) using parameters collected from prior studies and commercial datasheets~\cite{DNPU,dual-mode-cnn,hailo8_brief,VR-accelerator,micron_lpddr5x_datasheet,skhynix_lpddr5x_catalogue,PIM-AI,functino-in-memory,micron_automotive_ufs_datasheet,ufs2014}. In addition, the TECU operates on latent symbols, with throughput expressed in terms of the corresponding input frames. We synthesize the newly added PHY and TECU modules using Synopsys Design Compiler with the UMC 28nm technology library to estimate area and active power.

\begin{table}[t]
\centering
\caption{Area and power breakdown of the implemented cUFS, LPDDR-PIM, and TECU components (UMC 28nm).}
\label{area-breakdown}
\begin{tabular}{l l c c}
\toprule
\textbf{Module} & \textbf{Component} & \textbf{Area (mm$^2$)} & \textbf{Power (W)} \\
\midrule
\multirow{4}{*}{cUFS}
    & PE Array     & 0.82 & \multirow{3}{*}{0.42} \\
    & SRAM Buffers & 0.56 &                       \\
    & Vec Unit     & 0.16 &                       \\
\cmidrule(lr){2-4}
    & PHY          & 1.44 & 0.18                  \\
\midrule
\multicolumn{2}{c}{LPDDR-PIM} & 43.2 & 2.59 \\
\midrule
\multicolumn{2}{c}{TECU}      & 0.07 & 0.04 \\
\bottomrule
\end{tabular}
\end{table}

\textbf{Overall Power \& Area}. We use Synopsys Design Compiler to estimate the additional area of the NPU core, the TECU, and PHY in cUFS. For LPDDR-PIM, we estimate the chip area based on the reported die area of LPDDR5 DRAM and the area of the added near-bank compute units. Based on the energy/power models described above and the simulator configuration, we simulate workload execution and derive the average runtime power consumption. The resulting area and power estimates are summarized in~\autoref{area-breakdown}: cUFS and TECU introduce a total area overhead of 3.05 mm$^2$ and a power overhead of 0.64 W. The added area accounts for only a small fraction of the overall uMCP package area. 
The combined power of the cUFS, TECU, and PIM components is approximately 3.23 W, comparable to the approximately 3.27 W of near-Flash logic per package in \cite{lincoln}. The thermal feasibility reported for its vertically stacked memory–compute organization provides supporting evidence for the practicality of our integration approach.

\begin{table}[t]
\centering
\caption{Neural network compute and model size comparison under 2K resolution input.}
\small
\setlength{\tabcolsep}{4pt} %
\begin{tabular}{lccc}
\toprule
\textbf{Network} & \textbf{Enc.FLOPs (G)} & \textbf{Dec.FLOPs (G)} & \textbf{Params (M)} \\
\midrule
DsD-Image & 141.23 & 384.04 & 27.38 \\
DsD-Video & 215.04 & 437.55 & 35.51 \\
CDC-30    & 216.15 & 5270.23 & 35.79 \\
\bottomrule
\end{tabular}
\label{arch_workloads}
\end{table}

\section{Evaluation}

\begin{figure*}[t]
    \centering
    \includegraphics[width=0.95\textwidth]{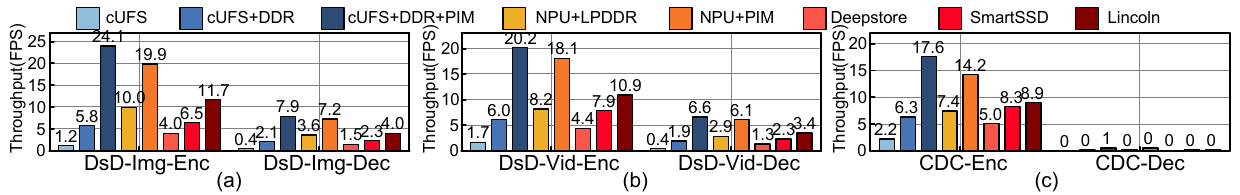}
    \caption{Performance of encoding and decoding stages under different architectures and workloads.}
    \label{arch-overall-fps}
\end{figure*}

\subsection{Implementation}
\textbf{Algorithm Level}. We implement Dual-stage Diffusion (DsD) for both image and video tasks. The models are trained on a server equipped with $2 \times$ NVIDIA A100 GPUs. The network is trained for 100 epochs with an initial learning rate of $1\times10^{-4}$ , and cosine decay is applied after 40 epochs. The training data is drawn from the Vimeo-90K\cite{vimeo90k} dataset.

We select diffusion codecs including CDC\cite{CDC}, DiffVC\cite{DiffVC}, and DiffEIC\cite{DiffEIC}, as well as traditional codecs such as JPEG, H.264, and H.265 as our algorithm baselines. We use the Tecnick and UVG 2K~\cite{uvg-dataset} datasets as our main benchmarks, as they have been widely adopted in neural codec studies. %
We mainly use PSNR, SSIM, LPIPS, and DISTS\cite{psnr_ssim,lpips,dists} as evaluation metrics, where LPIPS and DISTS are used to assess perceptual quality. We further report FID\cite{fid} as an auxiliary metric to evaluate the generative capability.

\textbf{cUFS Implementation}. To validate the effectiveness of our architecture, we simulate the performance of 5 architectures: (1) \textbf{NPU+DRAM (LPDDR)} as the baseline. (2) \textbf{cUFS}: Accessing SoC memory through the UFS controller. (3) \textbf{cUFS+DDR}: Accessing SoC memory via a single DDR controller enabled by the uMCP integration. (4) \textbf{cUFS+DDR+PIM}: Coordinated execution between cUFS and PIM. (5) \textbf{NPU+PIM (LPDDR)}: Collaborative execution between the NPU and PIM, used to compare against cUFS+PIM under matched compute capability. The reported performance and energy results include the initial weight-transfer cost, including flash–cUFS data movement.

In addition, we include \textbf{Deepstore}\cite{deepstore}, \textbf{SmartSSD}\cite{smartssd} and \textbf{Lincoln}\cite{lincoln} as near-data-processing baselines. We evaluate these baselines using workload-specific simulations in the same SCALESim-style framework, modeling their compute resources and data-transfer paths.
(1) \textbf{DeepStore}, which integrates compute units at the SSD, channel, and chip levels, allocating 4 TOPS to the UFS controller and 2 TOPS to each flash controller. The UFS side includes a 4 MB single-channel DRAM with 12.8 GB/s bandwidth, matching the original architecture, while SRAM size remains the same. Entropy coding and decoding execute on the host CPU, while neural-network operators are jointly executed by the UFS-controller and flash-controller compute units. (2) \textbf{SmartSSD}, which uses a PCIe-switched data path between flash and an external FPGA accelerator, where neural inference and entropy coding are performed. We model the FPGA with 12 TOPS peak compute, 19.2 GB/s internal memory bandwidth, and 3 GB/s data transfer bandwidth. (3) \textbf{Lincoln}, which uses a compute-enabled Flash architecture to enhance LLM inference performance. Each Flash die includes a Flash array and a logic layer with LPDDR PHY and near-Flash compute units. %
The Flash die has a read bandwidth of 8.48 GB/s, while the NPU maintains a performance of 4 TOPS. Near-Flash compute units peak at 9.6 TOPS, and the combined LPDDR bandwidth for DRAM and Flash ranks is 51.2 GB/s. Entropy coding and decoding execute on the host CPU, while neural-network operators are jointly executed by the host NPU and near-Flash compute units.

\begin{table}[t]
    \centering
    \caption{W8A8 QAT results on Tecnick.}
    \label{tab:quantization}
    \small
    \setlength{\tabcolsep}{3pt}
    \renewcommand{\arraystretch}{1.1}
    \begin{tabular*}{\columnwidth}
        {@{\extracolsep{\fill}}crrr@{}}
        \hline
        \textbf{\boldmath BPP ($\Delta$\%)}
        & \textbf{\boldmath $\Delta$PSNR}
        & \textbf{\boldmath $\Delta$MS-SSIM}
        & \textbf{\boldmath $\Delta$LPIPS} \\
        \hline
        .0650 ($-4.68$) & $+0.38$ & $+0.020$ & $+0.006$ \\
        .0885 ($-3.31$) & $+0.11$ & $+0.005$ & $+0.025$ \\
        .1028 ($-3.33$) & $-0.25$ & $\approx 0$ & $+0.046$ \\
        .1346 ($-2.56$) & $-0.44$ & $-0.001$ & $+0.041$ \\
        \hline
    \end{tabular*}
\end{table}

\begin{table}[t]
\centering
\caption{Ablation study of the global, local, and DsD components.}
\label{alg-ablation}
\footnotesize
\setlength{\tabcolsep}{2.8pt}
\renewcommand{\arraystretch}{1.05}
\begin{tabular}{c c c c c c c}
\toprule
\textbf{bpp} & \textbf{method}
& \textbf{PSNR$\uparrow$}
& \textbf{SSIM$\uparrow$}
& \textbf{LPIPS$\downarrow$}
& \textbf{DISTS$\downarrow$}
& \textbf{FID$\downarrow$} \\
\midrule

\multirow{3}{*}{0.06}
& global & 24.94 & 0.84  & 0.34  & 0.22  & 160 \\
& local  & -0.21 & -0.01 & 0.08  & 0.02  & 19  \\
& \textbf{DsD} & 0.08 & 0.00 & -0.03 & -0.01 & -12 \\
\midrule

\multirow{3}{*}{0.10}
& global & 30.08 & 0.92 & 0.21  & 0.13  & 68  \\
& local  & 0.05  & 0.00 & 0.04  & 0.01  & 2   \\
& \textbf{DsD} & 0.14 & 0.00 & -0.03 & -0.02 & -13 \\
\midrule

\multirow{3}{*}{0.32}
& global & 36.50 & 0.98 & 0.08  & 0.04  & 6  \\
& local  & 0.18  & 0.00 & 0.03  & 0.01  & 1  \\
& \textbf{DsD} & -0.23 & 0.00 & -0.02 & -0.01 & -2 \\
\midrule

\multicolumn{2}{c}{\textbf{Params \&  }}
& \textbf{global}
& \textbf{local}
& \textbf{DsD}
& \textbf{DiffEIC}
& \textbf{CDC} \\
\multicolumn{2}{c}{\textbf{latency}}
& 26M, 0.14s
& 23M, 0.12s
& 27M, 0.20s
& 1.4B, 82.5s
& 36M, 9.4s \\

\bottomrule
\end{tabular}
\end{table}

\subsection{Algorithm performance}

We evaluate DsD on Tecnick and UVG, as shown in \autoref{alg-perform}. Overall, DsD consistently achieves the best performance in terms of PSNR. For perceptual quality, DsD delivers comparable LPIPS/DISTS to other diffusion-based methods on both image and video workloads. The largest gap is observed between DsD and DiffEIC, where the LPIPS/DISTS difference is around $0.1$; such a gap is negligible in terms of visual quality. However, for generation-oriented metrics such as FID, DsD lags behind multi-step denoising-based generative codecs. This is mainly due to two reasons, first, the two-step diffusion process trades off part of the perceptual performance. Second, when the number of test samples ($< 1K$) is limited, FID can exhibit relatively large variance. 

To validate the effectiveness of DsD, we perform an ablation study on the Tecnick dataset. The term \textit{global} refers to full-resolution one-step diffusion, while \textit{local} indicates patchified diffusion. The \textit{local} and \textit{DsD} results show relative changes against \textit{global}. The findings reveal that \textit{local} degrades FID, LPIPS, and DISTS. Conversely, DsD mitigates this degradation and even achieves slight improvements, highlighting its ability to recover information lost during patchification. We also report model size and generation latency for a single 2K-resolution image on an RTX 4090 GPU. Results show that DsD strikes a favorable balance between perceptual quality and low latency.

\textbf{Quantization.}
Starting from FP16 checkpoints, we perform 1,000 W8A8 QAT
steps on random DIV2K crops with batch size 8.
Entropy-model computation and few activations retain
FP16, covering less than 5\% of operations.
Table~\ref{tab:quantization} reports Tecnick results: BPP is measured after QAT (relative change in
parentheses), and quality deltas are QAT minus FP16. Across these operating points, PSNR drops by at most
0.44~dB and LPIPS increases by at most 0.046.
At the lowest bitrate, PSNR and MS-SSIM improve despite
a 4.68\% bitrate reduction, with a small LPIPS increase of 0.006.

\begin{figure}[t]
    \centering
    \includegraphics[width=0.90\columnwidth]{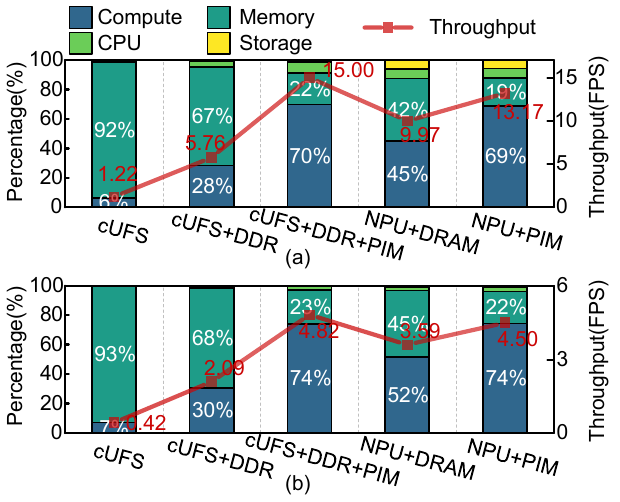}
    \caption{The throughput under 6-TOPS limitation and detailed time breakdown of (a)encoding and (b)decoding stage.}
    \label{arch-ablation-fps}
\end{figure}

\subsection{Architecture Performance}
As shown in \autoref{arch_workloads}, we select 3 workloads: DsD-image, DsD-video, and CDC(30 denoising steps) to evaluate end-to-end performance. We first evaluate the encoding and decoding throughput across all architectures, as shown in \autoref{arch-overall-fps}. The decoding throughput across different algorithms shows that \textit{Dual-stage Diffusion} achieves approximately $6\times$ higher performance than diffusion-based method (CDC). Across all the encoding and decoding tasks, our architecture achieves $2.42\times$ and $2.21\times$ average throughput improvements over the baseline, respectively. Compared with Deepstore, SmartSSD and Lincoln, cUFS+PIM delivers $4.99\times$, $3.02\times$ and $2.01\times$ higher decoding throughput, respectively.

\begin{figure}[t]
    \centering
    \includegraphics[width=0.90\columnwidth]{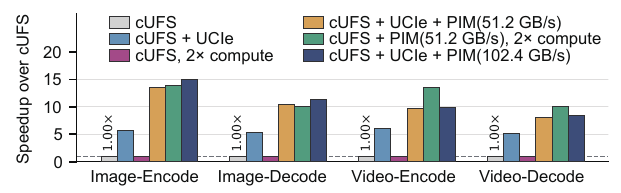}
    \caption{Architecture ablation study.}
    \label{arch-ablation}
\end{figure}

We evaluate the proposed architecture under aligned 6-TOPS total compute to obtain time breakdown. As shown in \autoref{arch-ablation-fps}, the main performance bottlenecks are computation and memory access. Our architecture achieves a performance improvement of $1.13\times$ for encoding and $1.07\times$ for decoding over the NPU+PIM, effectively reducing critical-path overhead from entropy coding and UFS data movement.

\begin{figure}[t]
    \centering
    \includegraphics[width=0.90\columnwidth]{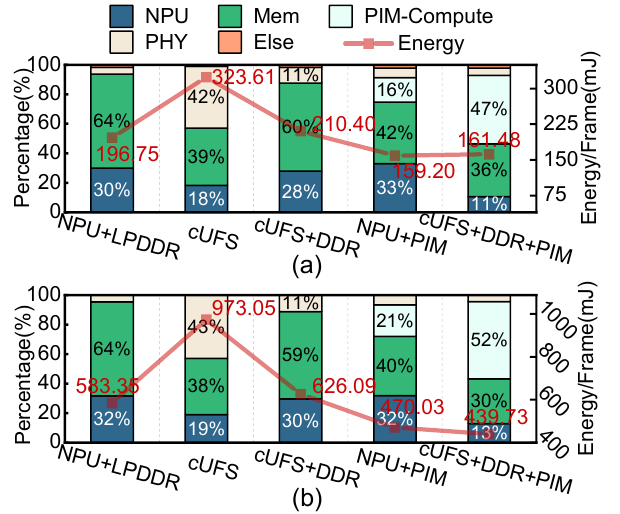}
    \caption{The Energy cost of one frame under 6-TOPS limitation and detailed energy breakdown of (a)encoding and (b)decoding stage.}
    \label{arch-ablation-energy}
\end{figure}

As shown in \autoref{arch-ablation-energy}, we breakdown the energy consumption into five components: NPU, memory accesses (including both internal and external accesses when PIM is used), PIM computation, PHY, and others.
The cUFS+PIM requires approximately $161.48 mJ$ for encoding and $439.73 mJ$ for decoding a single frame, achieving reductions of $17.93\%$ and $24.62\%$ over the NPU alone. In the cUFS+DDR+PIM design, PHY energy averages about $7.9 mJ$ for encoding and $19.7 mJ$ for decoding, reflecting reductions of $65.2\%$ and $71.7\%$ over cUFS+DDR. These results show that the collaboration between cUFS and PIM can minimize PHY I/O overhead.

\textbf{Architecture ablation.} \autoref{arch-ablation} compares cUFS and cUFS–PIM under a fixed 6-TOPS budget, allocated as 6/0 and 2/4 TOPS, respectively. We separately vary external bandwidth (w/o UCIe), PIM internal bandwidth (51.2 vs 102.4 GB/s), and compute capability. Adding UCIe improves cUFS throughput by up to $5.99\times$, whereas doubling compute alone yields approximately $1.00\times$, indicating a data-movement bottleneck. With identical UCIe connectivity and total compute, introducing PIM at 51.2 GB/s provides a further $1.58–2.35\times$ improvement. Doubling PIM internal bandwidth subsequently adds at most 10.9\%, indicating that the gains cannot be explained solely by the larger internal-bandwidth setting.

\begin{figure*}[t]
    \centering
    \includegraphics[width=0.95\textwidth]{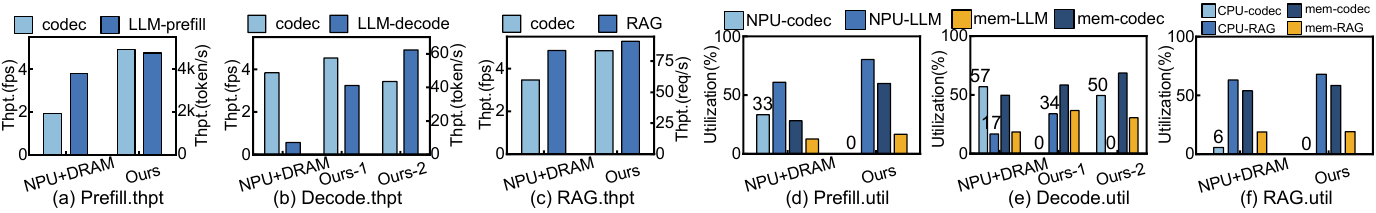}
    \caption{System performance under host-device co-run workloads.}
    \label{multi-workload}
\end{figure*}

\begin{figure}[t]
    \centering
    \includegraphics[width=0.90\columnwidth]{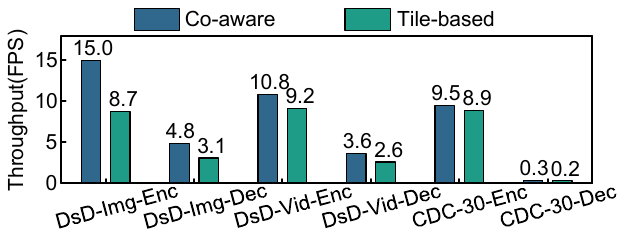}
    \caption{Performance improvements of patch-workload co-aware scheduling.}
    \label{exp_tiling}
\end{figure}

\begin{figure}[t]
    \centering
    \includegraphics[width=0.95\columnwidth]{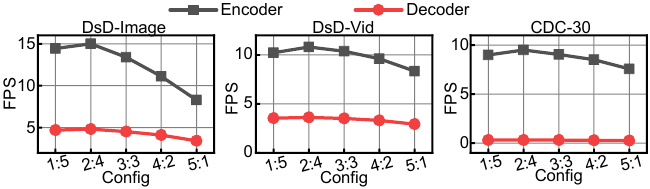}
    \caption{The impact of different compute allocations between cUFS and PIM on overall throughput.}
    \label{dse-tops}
\end{figure}

\subsection{Multi-workload Performance}
We evaluate end-to-end performance under three representative co-running AI workloads: Qwen-0.5B prefill with a 1,024-token sequence and decode with a 1,024-token prefilled KV cache, representing intensive accelerator usage, and Agent RAG with a 200-MB pre-built index, representing CPU-intensive execution. Agent RAG is built based on Haystack~\cite{haystack,cpu-centric} and BEAM~\cite{beam}. The diffusion decoder serves as the primary workload, while LLM prefill and Agent RAG run on the NPU and CPU, respectively. For LLM decode, we evaluate two exclusive mappings: NPU-only (Ours-1) and PIM-only (Ours-2). Codec execution follows our patch-workload-aware scheduling policy, while co-running neural-network workloads are scheduled at layer granularity. In both the baseline and proposed configurations, ready tasks are queued per execution device and dispatched in FCFS order when the device becomes available.

\autoref{multi-workload} (a)(d), (b)(e), and (c)(f) shows the throughput and resource utilization for LLM prefill, LLM decode, and Agent RAG, respectively. The NPU+DRAM architecture faces NPU contention between LLM and codec workloads, resulting in low utilizations of 61\% and 33\%. Switching to our proposed architecture allows the NPU to focus solely on the LLM workload, increasing its utilization to 80.2\%. This change leads to performance improvements of approximately 2.54$\times$ for the codec and 1.26$\times$ for the LLM workloads.

Offloading the decode stage of the LLM workload to the NPU boosts throughput by 5.56$\times$ for LLM decode and 1.18$\times$ for the codec. In contrast, offloading to PIM results in an 8.55$\times$ improvement for LLM decode, but decreases codec throughput by 0.4 FPS.
Our architecture enhances the performance of the Agent RAG workload by 1.03$\times$ and the codec workload by 1.39$\times$. Improvements for LLM workloads are modest since TECU contributes less to execution time compared to RAG. TECU primarily boosts bandwidth for the codec workload.

\subsection{Patch-Workload Co-Aware Scheduling}
We evaluate the patch-workload co-aware scheduling across six different workloads under 6TOPS limitation. As shown in \autoref{exp_tiling}, we observe an average throughput improvement of $1.34\times$, and up to $1.71\times$ on the DsD Encoding stage of image workloads. We further conduct DSE on compute resource allocation, as shown in \autoref{dse-tops}. Under a total compute budget of 6 TOPS, we vary the compute assigned to cUFS. The results show that the highest throughput is achieved when the cUFS compute is set to 2 TOPS and the cUFS:LPDDR-PIM is 1:2.

\section{Related Works}

\textbf{Algorithms}. Neural codecs have been studied~\cite{neural-img-compression-1, neural-img-compression-2, VIC-SH, DVC}. More recent diffusion-based codecs~\cite{CDC,HDCompression,DiffEIC,DiffVC,predictive-diffusion-1,predictive-diffusion-2} improve perceptual quality at extremely low bit rates, but largely overlook efficiency. Orthogonally, generic diffusion acceleration reduces denoising steps through sampler-based methods~\cite{ddim,dpmsolver} or training-based methods~\cite{distributionmatchingdistillation,Progressivedistillation,shortcut}, which approximate or learn shorter reverse trajectories from highly noisy states. In contrast, \textit{DiffusionSSD} exploits codec-specific structure, using the upsampled feature as an initialization over patchified inputs.

\noindent \textbf{Specialized Codec and Diffusion Accelerators.} Several works\cite{fpx-nic, Sun2025LearnedIC} optimize non-diffusion neural codecs through hardware-aware algorithm design, while EXION\cite{exion} exploits sparsity to accelerate diffusion. These works do not address the entropy-coding and storage-transfer stages of a codec. Our method instead targets storage-integrated execution and proposes a direct cUFS–PIM data path with dedicated entropy coding.

\noindent \textbf{Architecture and System}. Prior research has broadly investigated near-data compression. The \textit{compressed storage} integrates compression algorithms into storage devices. The most common approach\cite{compressed-ssd-lz4,compressed-ssd-survey,compressed-ssd-zdc} is to embed dedicated compression units within the SSD controller. Some prior works\cite{zns-csd,HA-CSD} explore Host-SSD collaborative compression, which offloads computation to the host. Prior designs target conventional lossy or lossless compression, whose capability is insufficient.  Systems such as FastServe\cite{fastserve} and AlpaServe\cite{alphaserve} optimize request scheduling and multi-GPU model placement to improve resource efficiency. DiffusionSSD addresses resource contention  on edge devices by moving codec computation and entropy coding into the storage–memory subsystem, which is a different problem.

\section{Conclusion}

We introduce \textbf{DiffusionSSD}, a diffusion codec system for visual data compression on edge platforms. %
This is the first end-to-end deployment solution for generative codec workloads through coordinated algorithm, architecture, and system co-design. Experimental results show a 2.32× improvement in coding throughput over current solutions, while maintaining similar compression quality to existing diffusion codecs.

\bibliographystyle{IEEEtranS}
\bibliography{diffusionssd}

\end{document}